\documentclass[letterpaper]{article} 
\usepackage{aaai2026}  
\usepackage{times}  
\usepackage{helvet}  
\usepackage{courier}  
\usepackage[hyphens]{url}  
\usepackage{graphicx} 
\usepackage{natbib}  
\usepackage{caption} 
\usepackage{algorithm}
\usepackage{algorithmic}
\usepackage{booktabs}
\usepackage{amsmath}
\usepackage{graphicx}  
\usepackage{caption}    
\usepackage{subcaption} 

\usepackage{booktabs}   
\usepackage{multirow}  
\usepackage{array}     
\usepackage{amsfonts}
\usepackage{listings}

\DeclareCaptionType{listing}[Listing][List of Listings]
\usepackage[most,skins,theorems]{tcolorbox}

\usepackage{newfloat}
\usepackage{listings}
\DeclareCaptionStyle{ruled}{labelfont=normalfont,labelsep=colon,strut=off} 
\floatstyle{ruled}
\newfloat{listing}{tb}{lst}{}
\floatname{listing}{Listing}
\title{Beyond Detection: A Comprehensive Benchmark and Study on Representation Learning for Fine-Grained Webshell Family Classification}

\author {
    Feijiang Han
}
\affiliations {
    Department of Computer and Information Science, University of Pennsylvania\\
    Philadelphia, PA, USA\\
    feijhan@seas.upenn.edu
}

\usepackage{float}

\begin{document}

\maketitle

\begin{abstract}
Malicious WebShells pose a significant and evolving threat by compromising critical digital infrastructures and endangering public services in sectors such as healthcare and finance. While the research community has made significant progress in WebShell detection (i.e., distinguishing malicious samples from benign ones), we argue that it is time to transition from passive detection to in-depth analysis and proactive defense. One promising direction is the automation of WebShell family classification, which involves identifying the specific malware lineage in order to understand an adversary's tactics and enable a precise, rapid response.
This crucial task, however, remains a largely unexplored area that currently relies on slow, manual expert analysis. To address this gap, we present the first systematic study to automate WebShell family classification. Our method begins with extracting dynamic function call traces to capture inherent behaviors that are resistant to common encryption and obfuscation. To enhance the scale and diversity of our dataset for a more stable evaluation, we augment these real-world traces with new variants synthesized by Large Language Models. These augmented traces are then abstracted into sequences, graphs, and trees, providing a foundation to benchmark a comprehensive suite of representation methods. Our evaluation spans classic sequence-based embeddings (CBOW, GloVe), transformers (BERT, SimCSE), and a range of structure-aware algorithms, including Graph Kernels, Graph Edit Distance, Graph2Vec, and various Graph Neural Networks. Through extensive experiments on four real-world, family-annotated datasets under both supervised and unsupervised settings, we establish a robust baseline and provide practical insights into the most effective combinations of data abstractions, representation models, and learning paradigms for this challenge.
This foundational work is a crucial step toward automating threat intelligence, accelerating incident response, and ultimately enhancing the resilience of the digital services that society depends on.
\end{abstract}


\section{Introduction}
\label{sec:introduction}

Malicious WebShells have evolved from simple scripts into strategic assets used in sophisticated attacks that directly threaten critical public services in sectors like healthcare and finance, endangering the sensitive data of millions. To counter this pervasive threat, the research community has achieved considerable success in developing automated techniques for WebShell detection \cite{tu2014WebShell,aboaoja2022malware,ma2024research, feng2024glareshell, han2025can}.

While successful, this focus on binary classification (malicious vs. benign) provides only a foundational first line of defense and offers limited actionable intelligence for subsequent security operations. A more proactive and robust security posture requires not just knowing that a server is compromised, but understanding the specific nature of the threat itself. This necessitates \textbf{WebShell family classification}: the task of identifying the specific variant or lineage of the malware. Automating this process is crucial as it unlocks a deeper level of threat intelligence, helping security teams attribute attacks, anticipate an adversary's next moves, and mount a faster, more targeted incident response \cite{zhao2024malicious}. For instance, an automated system can reduce response time from hours of manual expert analysis to mere seconds, enabling security operation centers (SOCs) to trigger specific defense playbooks tailored to a family's known tactics before significant damage, like data exfiltration, occurs. This critical task, however, remains largely unexplored in the research community, with current practices relying on time-consuming manual analysis.

We argue that automating this task is technically feasible for two primary reasons. First, WebShells within the same family often share distinct behavioral characteristics due to code reuse \cite{wrench2015towards, starov2016no}. Second, this malicious behavior can be captured in the program's dynamic function call traces even when the source code is obfuscated \cite{de2018now,xu2023family}. This insight forms our core hypothesis: by learning to recognize these fundamental behavioral patterns, a model can effectively group and track WebShell families, even when they are protected by surface-level obfuscation.

However, family classification is inherently more challenging than binary detection, as it requires models that can capture the nuanced behavioral patterns that differentiate families, not just generic malicious traits. This challenge motivates the foundational research question of our work: \textbf{\textit{What data structures and representation methods are most effective for capturing these family-specific behaviors?}}

To answer this question, this paper presents the first systematic study to benchmark WebShell family classification. We conduct a large-scale empirical evaluation of diverse data abstractions and representation learning methods designed to capture WebShell behavior. Our goal is to establish a robust foundation and a practical guide for this critical task.

Our contributions are as follows:
\begin{itemize}
    \item \textbf{A Comprehensive Methodological Framework.} We design and execute the first large-scale benchmark for this task. To ensure a robust evaluation, we introduce a data synthesis framework leveraging a Large Language Model (LLM) to augment our real-world data with diverse, behaviorally-consistent function call traces. We abstract this enriched dataset into three fundamental data structures (sequences, graphs, and trees) and systematically evaluate a diverse spectrum of representation learning methods, from classic word embeddings and transformers to structure-aware algorithms like Graph Kernels and various Graph Neural Networks (GNNs).

    \item \textbf{A Robust Empirical Baseline.} Through extensive experiments on four real-world datasets with both supervised and unsupervised classification, we establish the first robust, data-driven performance baseline for WebShell family classification. This provides a crucial point of comparison for all future work in this emerging area.

    \item \textbf{Actionable Insights for the Security Community.} Our analysis delivers a clear hierarchy of performance, demonstrating that structural representations (especially trees) are decisively more effective than sequential ones, and that GNNs are the premier modeling architecture. These findings offer immediate, practical guidance for practitioners and researchers aiming to build effective classification systems.

    \item \textbf{A Practical Guide to Implementation.} We distill our findings into a set of best practices for implementation, detailing optimal strategies for model selection and hyperparameter configuration. 
\end{itemize}

Ultimately, this work provides both a foundational benchmark and a practical guide, empowering the community to move beyond simple detection and build the next generation of intelligent, fine-grained defense systems.

\section{Problem Formulation}
\label{sec:problem_formulation}

The primary goal of WebShell family classification is to automatically categorize a given malicious WebShell into one of several predefined families. Our central research objective is to fundamentally understand which data abstractions and representation methods are most effective at capturing these family-specific features from raw data. We therefore adopt a two-stage framework that decouples representation learning from classification, enabling a fair and standardized benchmark of different encoders.

\paragraph{Stage 1: Representation Learning.}
The input to this stage is a raw, unstructured function call trace from a single WebShell. An encoder model, $g$, maps this trace into a fixed-dimensional numerical vector $\mathbf{x} = g(\text{trace}) \in \mathbb{R}^d$.

This vector $\mathbf{x}$, or embedding, is a structured summary of the WebShell's runtime behavior. Our core investigation lies in comparing various designs for this encoder $g$.

\paragraph{Stage 2: Benchmarking via Classification.}
In this stage, we use a suite of standard classifiers to benchmark the quality of the embeddings ($\mathbf{x}$) produced in Stage 1. The central principle is that a higher-quality representation will be more separable in vector space and thus yield better performance on classification tasks, providing an objective measure of the encoder's effectiveness. Formally, the classification task is defined as follows: given a dataset of embeddings $\mathcal{D} = \{(\mathbf{x}_1, y_1), \dots, (\mathbf{x}_n, y_n)\}$, where $y_i \in \{1, \dots, K\}$ is the family label for one of $K$ families. The objective is to learn a classifier $f: \mathbb{R}^d \to \{1, \dots, K\}$ that accurately predicts the label $\hat{y} = f(\mathbf{x})$ for any given embedding $\mathbf{x}$.

\section{Dataset Collection}
\label{sec:data_collection}

\paragraph{Data Acquisition and Annotation.}
Our dataset construction follows an established pipeline for creating high-quality, real-world WebShell datasets~\cite{zhao2024malicious}. The process begins with suspicious files flagged by a large-scale cloud provider's malware detection system. Each potential WebShell is executed in a secure sandbox to capture its dynamic function call trace---a chronologically ordered log of its runtime behavior.
This dynamic approach offers a significant advantage over static analysis, bypassing common evasion techniques like obfuscation and encryption, thereby revealing a well-defined structure of operational behaviors ideal for extracting family-specific features. 
Besides, this trace is essentially language-agnostic. By discarding noise from programming syntax and idiosyncratic coding habits, it focuses on the core operational logic, which makes our evaluation broadly generalizable to different server-side languages.
An example of this raw data is shown in Table~\ref{tab:dynamic_calls}.

\begin{table}[h!]
\centering
\caption{An Example of raw dynamic function call trace captured from sandboxed execution. Each record consists of a unique identifier and the corresponding function calls.}
\label{tab:dynamic_calls}
\begin{tabular}{@{}>{\raggedright\arraybackslash}m{2cm}>{\raggedright\arraybackslash}p{5.6cm}@{}}
\toprule
\textbf{Filemd5} & \textbf{Dynamic Function Calls} \\ \midrule
191c2...3b01 & [\textit{main}, \textit{zend\_compile\_file}, \textit{main}, \textit{base64\_decode}, \textit{main}, \textit{assert}, \textit{assert}, \textit{zend\_compile\_string}, \textit{assert}, \textit{zend\_fetch\_r\_post}, \textit{assert}, \textit{eval}, \textit{eval}, \textit{zend\_compile\_string}, \ldots] \\ 
\bottomrule
\end{tabular}
\end{table}

Security experts then manually review these traces to filter out false positives. The verified malicious samples subsequently undergo a human-machine collaborative process for family annotation. Samples that cannot be confidently assigned to any established family are designated as outliers and assigned a `Family ID` of -1. The final labeled data format is presented in Table~\ref{tab:family_labels}.

\begin{table}[h!]
\centering
\caption{Examples of annotated samples with their identifiers and assigned family labels.}
\label{tab:family_labels}
\begin{tabular}{@{}lc@{}}
\toprule
\textbf{Filemd5} & \textbf{Family ID} \\ \midrule
12b7340d1b8acf0fe2d78fce84bccf8c & 1 \\
1aba8701dcab6629caa9e21fc772b50e & 2 \\
28c5678442c6a3ee17290ece4d1c8904 & 3 \\
00cd0f1bfda4903dba26541301c686ec & 5 \\
01625e53cb2d1275fbf4b2af0f6946e3 & -1 \\ \bottomrule
\end{tabular}
\end{table}

\paragraph{LLM-Powered Data Augmentation.}

To address the inherent limitations of real-world data collection, such as data scarcity for rare families and the absence of novel, zero-day threats, we introduce an LLM-augmented data synthesis framework to enrich and expand our dataset. This framework enables us to scale our data collection efforts and enhance the diversity of the samples. The specific prompt templates used in this process are detailed in the Appendix.

Specifically, we employed a two-pronged strategy. First, for \textbf{Intra-Family Data Augmentation}, we used a few-shot prompts, providing the LLM with a high-level description of a family's behavior along with several canonical examples of its function call traces. This enabled the model to generate a large volume of new samples that are behaviorally consistent with existing families but syntactically unique. This technique effectively addresses the class imbalance problem by augmenting underrepresented families. 

Second, building on this, we introduced a \textbf{New Family \& Zero-Day Simulation}. This stage simulates the adversarial tactic of creating novel variants by blending the behavioral characteristics of different malware families. The resulting synthetic traces can be labeled either as entirely new families or as adversarial outliers, which are designed to challenge the classifier's robustness.

To ensure the fidelity and logical soundness of the augmented synthetic data, all LLM-generated traces underwent a rigorous two-stage, human-in-the-loop verification and sanitization process.

\begin{enumerate}
\item \textbf{Automated Filtering:} All generated traces first pass through an automated filter to ensure they are well-formed (e.g., proper list format and valid function names from our vocabulary). Malformed traces are immediately discarded.
\item \textbf{Human-in-the-Loop Verification:} We then use a visualization platform to project and plot the embeddings of the filtered synthetic samples alongside the real-world samples from the same family. For intra-family data augmentation samples, we manually review these clusters and discard any synthetic samples that significantly deviate from their family's core cluster center.
\end{enumerate}

\paragraph{Dataset Details.} 
Through this process, we constructed four distinct datasets for our experiments, labeled DS1 through DS4. These datasets feature progressively increasing scale and complexity in terms of sample size, the number of families, and the quantity of outliers. This graduated design allows for a robust and thorough evaluation of our representation methods across diverse conditions. Table~\ref{tab:dataset_details} summarizes the key statistics of each dataset.

\begin{table}[h!]
\centering
\caption{Details of our dataset. The complexity increases from DS1 to DS4.}
\label{tab:dataset_details}
\resizebox{\columnwidth}{!}{%
\begin{tabular}{@{}lcccc@{}}
\toprule
\textbf{Dataset} & \textbf{\# Samples} & \textbf{Complexity} & \textbf{\# Families} & \textbf{\# Outliers} \\ \midrule
DS1              & 452                 & Low                 & 21                   & 1                    \\
DS2              & 553                 & Medium              & 37                   & 10                    \\
DS3              & 1125                & High                & 48                   & 23                   \\
DS4              & 1617                & High                & 81                   & 28                   \\ \bottomrule
\end{tabular}
}
\end{table}

\begin{figure*}[h!]
    \centering
    \begin{subfigure}[b]{0.32\textwidth}
        \centering
        \includegraphics[width=\textwidth]{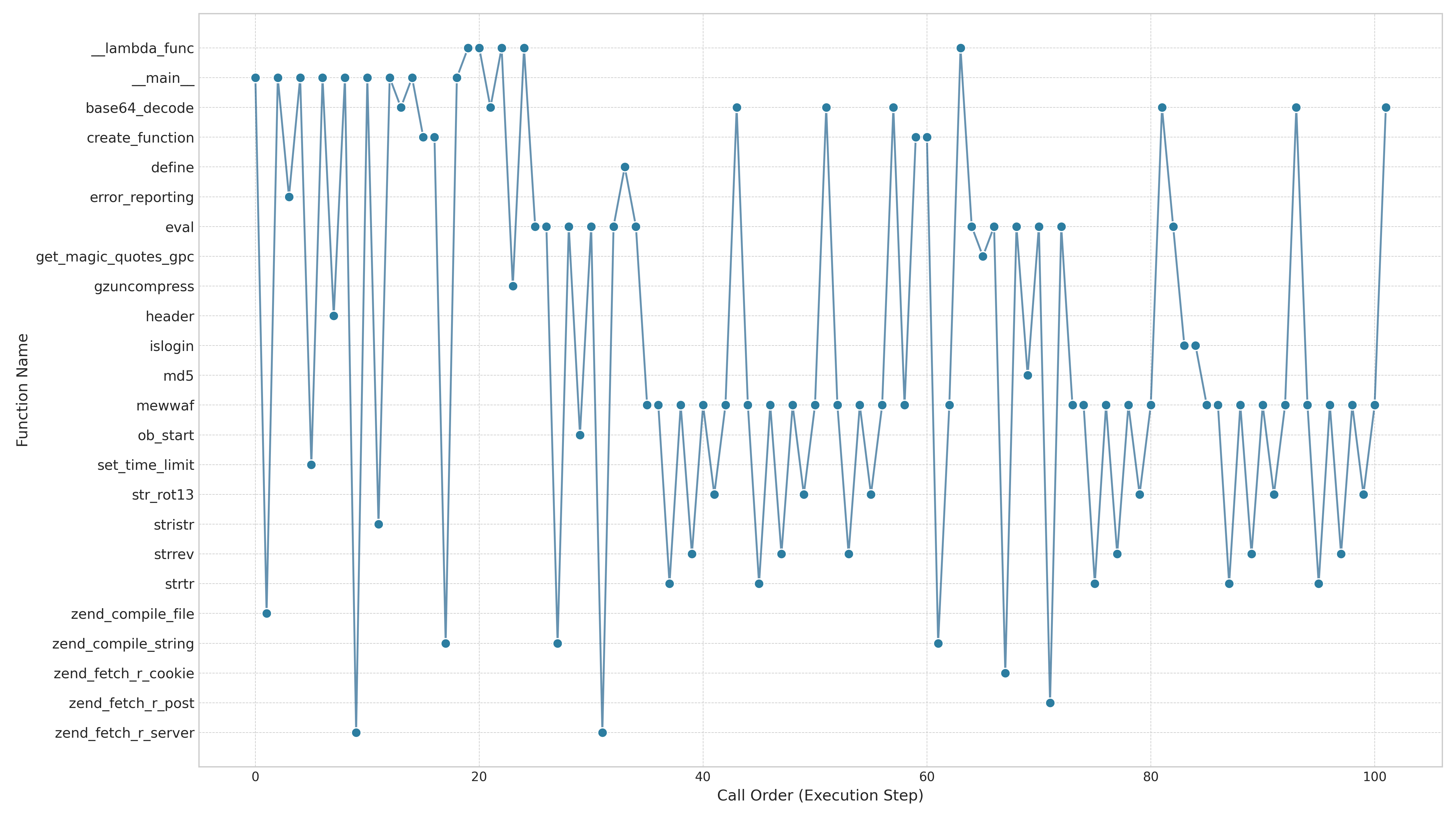}
        \caption{Sequence Model}
        \label{fig:seq_model}
    \end{subfigure}
    \hfill
    \begin{subfigure}[b]{0.32\textwidth}
        \centering
        \includegraphics[width=\textwidth]{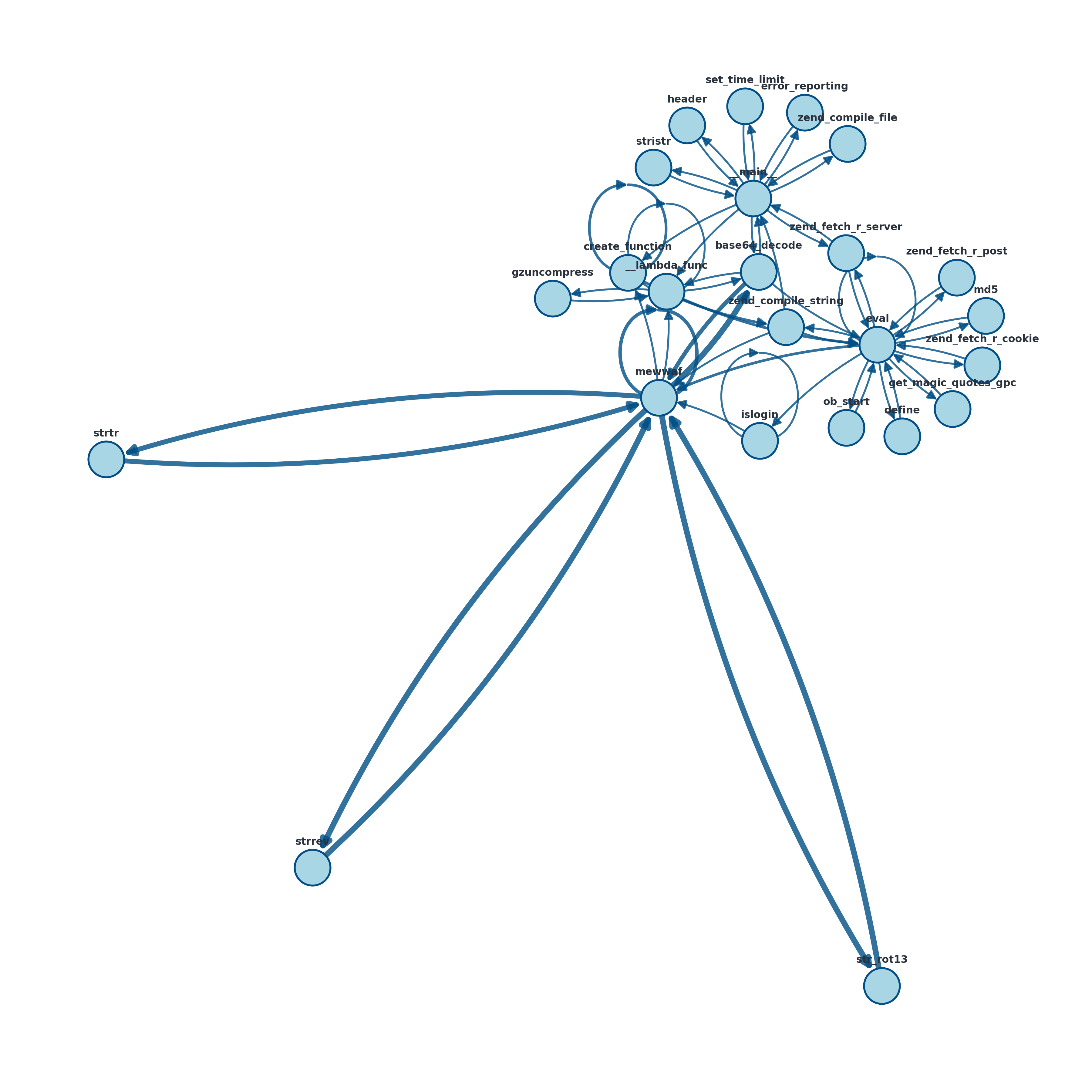}
        \caption{Graph Model}
        \label{fig:graph_model}
    \end{subfigure}
    \hfill
    \begin{subfigure}[b]{0.32\textwidth}
        \centering
        \includegraphics[width=\textwidth]{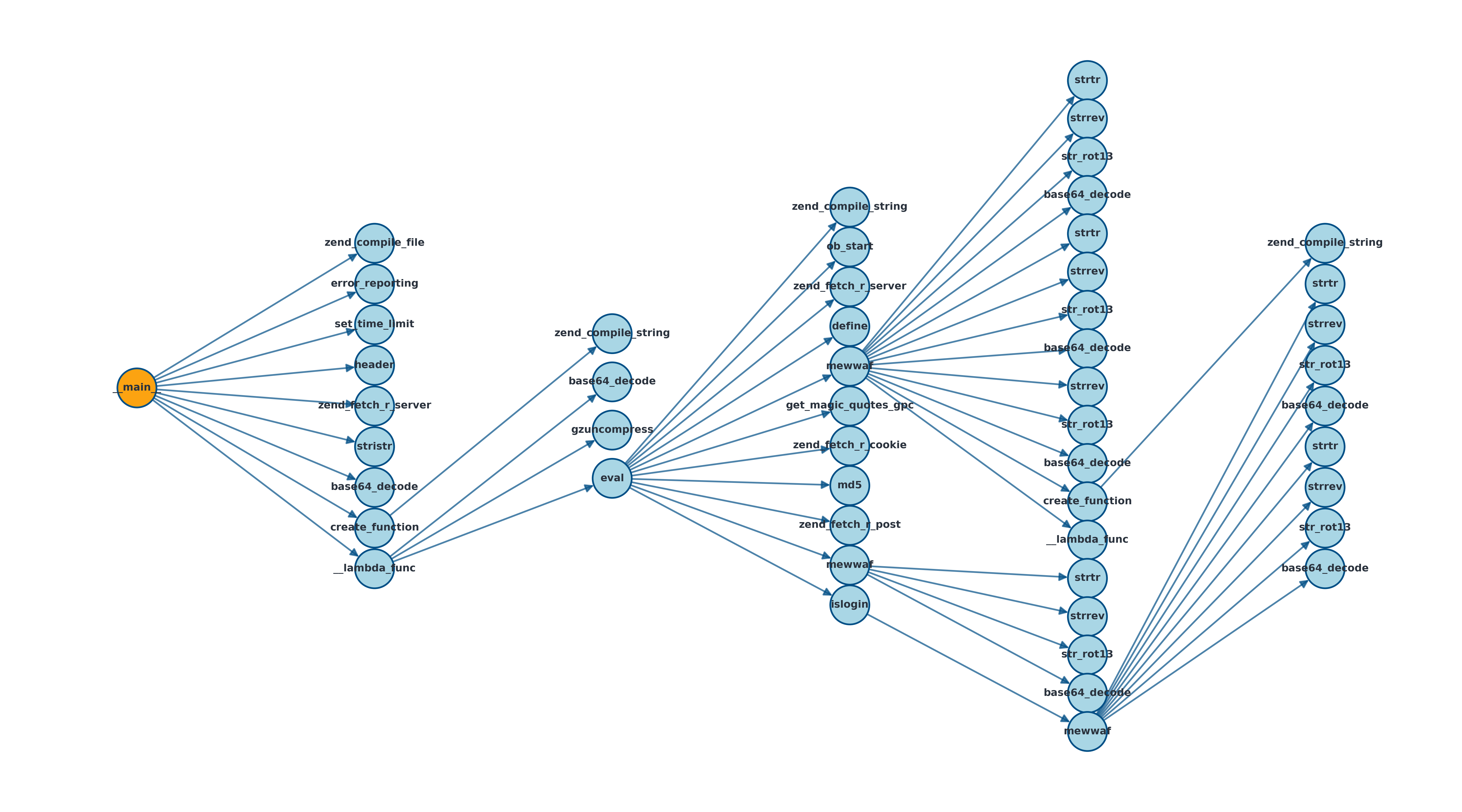}
        \caption{Tree Model}
        \label{fig:tree_model}
    \end{subfigure}

    \caption{%
        The visualization of three data abstractions.
        (a) The Sequence Model visualizes the chronological execution flow.
        (b) The Graph Model provides a static, aggregate view of all calling relationships.
        (c) The Tree Model preserves the hierarchical call structure and execution context.
    }
    \label{fig:data_abstractions}
\end{figure*}

\section{Behavioral Data Abstraction}
\label{sec:data_abstraction}

To make raw function call traces amenable to machine learning, we abstract this sequential data into three distinct structural representations: sequences, graphs, and trees. As illustrated in Figure~\ref{fig:data_abstractions}, each representation captures a different aspect of a WebShell's runtime behavior, providing a unique lens through which to analyze its characteristics.

\paragraph{Sequence Model.} The most direct abstraction treats a function call trace as a sequence of discrete tokens, where each function name becomes a token in the execution order (Figure~\ref{fig:seq_model}). This linear representation is compatible with a wide range of natural language processing models. A trace can be represented as $S = (t_1, t_2, \dots, t_n)$, where $t_i$ is the $i$-th function called.

\paragraph{Graph Model.} To capture more complex, non-sequential interactions, we model each trace as a Function Call Graph (FCG), shown in Figure~\ref{fig:graph_model}. An FCG, $G = (V, E)$, provides a static, aggregate view of the program's behavior, where each unique function is a node $v \in V$, and a directed edge $(u, v) \in E$ exists if function $u$ ever calls function $v$. The edges can be weighted by call frequency to represent the strength of the interaction. This model effectively captures all calling relationships, including loops and indirect calls.

\paragraph{Tree Model.} To preserve the hierarchical nature of program execution, we also represent each trace as a Function Call Tree (FCT), illustrated in Figure~\ref{fig:tree_model}. The FCT, $T = (V, E)$, is a rooted tree where the entry point (e.g., `\_main\_`) is the root and edges represent direct parent-child call relationships. Unlike the graph model, the FCT is acyclic and preserves the specific execution path and context; a function called multiple times in different contexts appears as distinct nodes in the tree.

\section{Representation and Benchmarking}
\label{sec:representation_methods}

\subsection{Representation Learning Methods}
We apply a diverse set of foundational and widely adopted representation learning techniques tailored to each data abstraction. Table~\ref{tab:methods_summary_full} provides a complete overview.

\paragraph{For sequence models,} we evaluate two distinct categories: classic context-free methods (CBOW~\cite{mikolov2013efficient}, GloVe~\cite{pennington2014glove}) and modern context-aware transformers (BERT~\cite{devlin2018bert}, SimCSE~\cite{gao2021simcse}). To produce a single fixed-dimensional vector for each function call trace, we employ several aggregation strategies tailored to each model type. For the static embeddings produced by CBOW and GloVe, we investigate three strategies: averaging all function call vectors to create a mean representation (avg); concatenating the vectors of each function call sequentially (concat); and a TF-IDF weighted average that emphasizes more discriminative functions. For the transformer models, we leverage their deep, contextualized hidden states by: averaging the hidden states across all tokens (avg); concatenating the hidden state vectors from different layers (concat); and using the final hidden state of the dedicated classification token, [CLS].

\paragraph{For graph and tree models,} we employ two classic methods for direct structural comparison. First, Graph/Tree Kernels~\cite{shervashidze2011weisfeiler} measure similarity by counting shared substructures, such as common call sequences (paths), randomly generated traversals (random walks), and identical small-scale call hierarchies (subtrees). Second, we compute the Graph/Tree Edit Distance~\cite{marzal1993computation}, which quantifies dissimilarity by calculating the minimum cost of operations (e.g., node insertion, deletion, and substitution) required to transform one structure into another. For learning-based approaches, we benchmark three prominent Graph Neural Network (GNN) architectures which learn representations via message passing: Graph Convolutional Network (GCN)~\cite{kipf2016semi}, Graph Attention Network (GAT)~\cite{velickovic2018graph}, and Graph Isomorphism Network (GIN)~\cite{xu2018how}. Finally, we include Graph2Vec~\cite{narayanan2017graph2vec}, an unsupervised method that learns whole-graph embeddings.

\begin{table}[h!]
\centering
\caption{Overview of the representation methods and their implementation variants evaluated for each data abstraction.}
\label{tab:methods_summary_full}
\resizebox{\columnwidth}{!}{%
\begin{tabular}{@{}ll@{}}
\toprule
\textbf{Representation Method} & \textbf{Implementation Variants} \\ \midrule
\multicolumn{2}{@{}l}{\textit{\textbf{Sequence-Based Models}}} \\
\cmidrule(l){1-2}
Word2Vec (CBOW) & Concat, Avg, Concat \& Avg \\
GloVe & Concat, Avg, Concat \& Avg \\
BERT & Concat, Avg, CLS \\
SimCSE & Concat, Avg, CLS \\
\midrule
\multicolumn{2}{@{}l}{\textit{\textbf{Graph-Based Models}}} \\
\cmidrule(l){1-2}
Graph Kernel & Path, Walk, Subtree \\
Graph Edit Distance & -- \\
Graph Neural Networks & GCN, GAT, GIN \\
Graph Embedding (Graph2Vec) & -- \\
\midrule
\multicolumn{2}{@{}l}{\textit{\textbf{Tree-Based Models}}} \\
\cmidrule(l){1-2}
Tree Kernel & Path, Walk, Subtree \\
Tree Edit Distance & -- \\
Graph Neural Networks & GCN, GAT, GIN \\
Tree Embedding (Graph2Vec) & -- \\ \bottomrule
\end{tabular}
}
\end{table}

\paragraph{Benchmarking Classifiers.}

To assess the quality of the learned representations, we benchmark the resulting embeddings using a suite of four standard classifiers. For unsupervised evaluation, we use K-Means~\cite{macqueen1967some} and Mean-Shift~\cite{comaniciu2002mean} clustering. For supervised evaluation, we employ two widely-used models: Random Forest~\cite{breiman2001random} and the Support Vector Machine (SVM)~\cite{cortes1995support}. This comprehensive framework allows us to measure the effectiveness of each representation in both labeled and unlabeled settings.

\section{Experimental Setup}

\begin{figure*}[h!]
    \centering
    \includegraphics[width=0.95\linewidth]{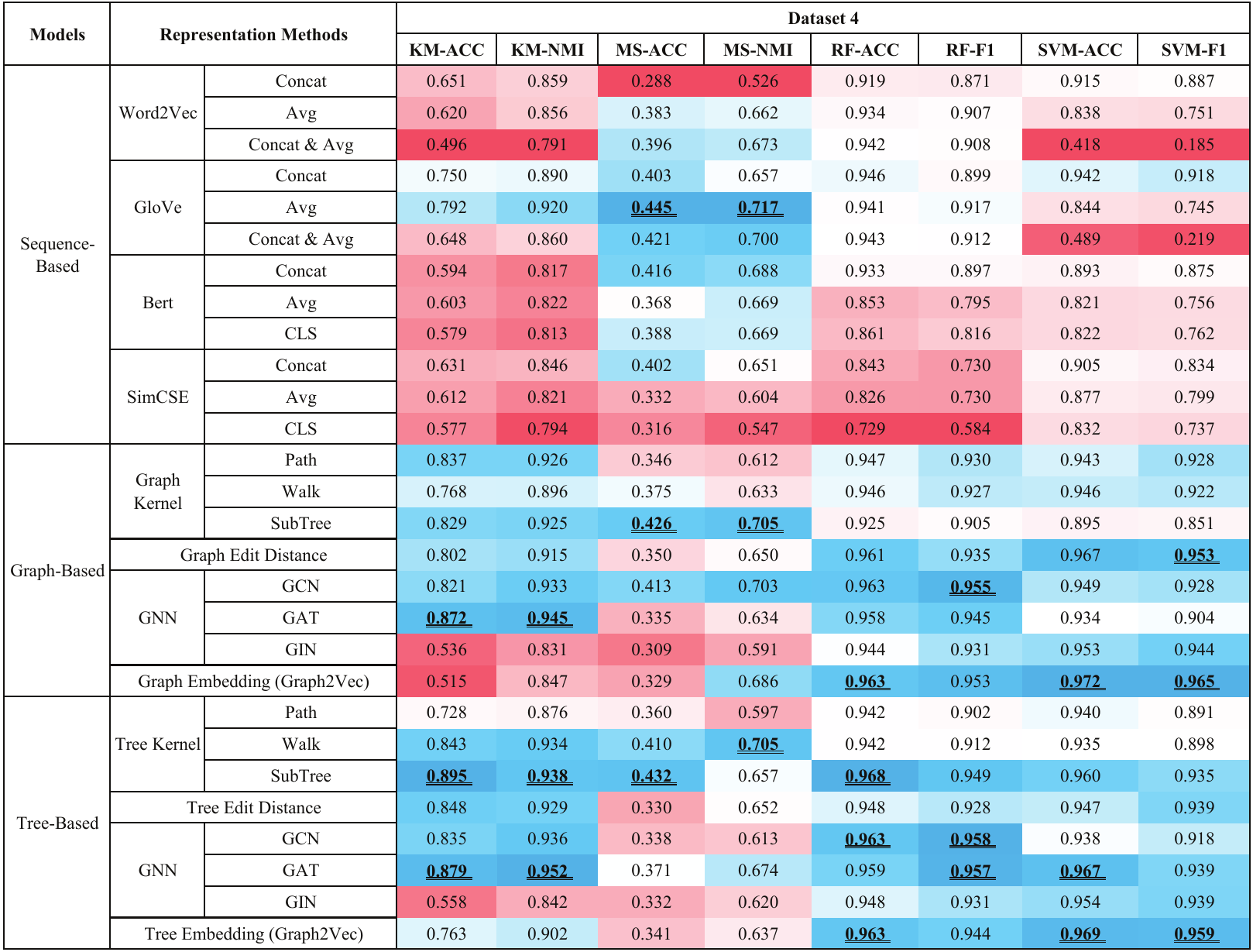}
    \caption{Performance comparison of representation methods on the \textit{DS4 dataset}. Columns denote classifiers (KM: K-Means; MS: Mean-Shift; RF: Random Forest; SVM) and metrics.}
    \label{fig:ds4}
\end{figure*}

\subsection{Implementation Details}

\paragraph{Representation Models.}
For all representation learning models, we standardized the output embedding dimension to 128 to balance expressiveness and computational efficiency.
The input dimensions were dynamically set based on the function vocabulary size of each specific dataset. To establish a consistent and reproducible baseline, we utilized the default hyperparameter settings (e.g., optimizer, learning rate, loss function) recommended for each model during the representation learning phase. Detailed configurations for each model are provided in the Appendix.

\paragraph{Benchmarking Classifiers.}
To ensure a fair comparison across different representations, we employed a grid search with cross-validation to tune the hyperparameters for each representation-classifier pair. To ensure statistical robustness, all reported results are the average of 10 independent runs, each with a different seed.

\subsection{Evaluation Metrics}

\paragraph{Supervised Classification.}
We assess the performance of supervised models using standard metrics: Accuracy and F1-score. For the multi-class setting, F1-score is reported as macro-averaged values. This approach computes the metric independently for each family and then calculates the unweighted mean, ensuring that all families, regardless of their size, contribute equally to the final score.

\paragraph{Unsupervised Clustering.}
We evaluate clustering quality using two primary metrics: Accuracy and Normalized Mutual Information (NMI). Accuracy is computed by first finding the optimal mapping between cluster assignments and ground-truth labels via the Hungarian algorithm and then calculating the percentage of correctly assigned samples. NMI measures the agreement between the assigned clusters $C$ and true labels $Y$, correcting for chance:
\begin{equation}
    \operatorname{NMI}(Y, C) = \frac{2 \times I(Y; C)}{H(Y) + H(C)},
\end{equation}
where $I(Y; C)$ is the mutual information between the true and predicted labels, while $H(Y)$ and $H(C)$ are their respective entropies.

\section{Results and Analysis}
\label{sec:analysis}

We present our experimental findings for the most complex \textbf{DS4 datasets} in Figures~\ref{fig:ds4}. The full results for DS1, DS2, and DS3 are in the Appendix.
Performance is visualized using a blue-to-red color gradient, where blue signifies higher scores. 
The top results in each column are highlighted. 

\subsection{Key Insight 1: Structural Semantics Definitively Outperform Sequential Syntax}

The most striking result from our benchmark is the significant performance gap between structural (graph and tree) and sequential representations. As shown in Figures~\ref{fig:ds4}, GNNs and even classic methods like Tree Edit Distance consistently achieve F1-scores exceeding 0.9, while the performance of advanced sequence models like BERT is both lower and more volatile. Notably, as dataset complexity increases, structural methods exhibit a much more graceful performance degradation, reinforcing their \textbf{robustness} and \textbf{scalability} for this task.

This performance delta highlights a fundamental limitation of sequential models. They are designed to capture linear dependencies, treating a function call trace as a syntactic sentence. However, a WebShell family's signature lies not in call adjacency but in the overarching control-flow topology. Malicious actors frequently reuse a core set of utility functions (e.g., for execution, encoding, or file access) but invoke them from different program locations, often inserting non-functional "junk" calls to thwart simple pattern matching.

Structural representations, by abstracting the trace into a graph or tree, capture these complex, non-local relationships. They model the program's \textit{who-calls-whom} relationships which is a far more fundamental and stable indicator of shared malicious logic. This inherent robustness to superficial code reordering and obfuscation is the primary reason for their superior performance.

\begin{table}[h!]
\centering
\caption{The top 3 representation methods for each classifier, ranked by overall performance across all datasets.}
\label{tab:optimal_methods_by_classifier}
\begin{tabular}{@{}ll@{}}
\toprule
\textbf{Classifier} & \textbf{Top 3 Optimal Representation Methods} \\ \midrule
K-Means & Tree-GAT, Graph-GAT, Tree-Kernel \\
Mean Shift & Tree-GAT, CBOW, GloVe \\
Random Forest & Tree-GCN, Graph-GCN, Tree-GAT \\
SVM & Tree-GAT, Graph-GIN, Tree-GIN \\ \bottomrule
\end{tabular}
\end{table}

\subsection{Key Insight 2: Hierarchical Context is Crucial, Granting Trees an Edge}

While both graph and tree models prove effective, our results show that tree-based representations yield better overall performance (Table~\ref{tab:optimal_methods_by_classifier}). This underscores the critical importance of \textbf{hierarchical context} in distinguishing WebShell families.

The reason for this advantage lies in what each structure preserves. A standard FCG is an aggregate view, merging all invocations of a function into a single node. It shows that function $A()$ called $B()$, but loses the context of how that call occurred. In contrast, a FCT is acyclic and preserves the precise execution path. Each node in an FCT represents a unique function invocation within a specific call stack.

This distinction is vital. A polymorphic function like $eval()$ might be used for different purposes depending on its caller. An FCT disambiguates these cases, representing $handler\_1() \rightarrow eval()$ and $handler\_2() \rightarrow eval()$ as distinct nodes with different parent-child relationships. This fine-grained, contextual fingerprint provides a more potent feature set for learning.

\subsection{Key Insight 3: GNNs are the Premier Architecture for Learning Behavioral Topologies}

Among all models, GNNs emerge as the most powerful and stable architecture, particularly the GAT and GCN. The theoretical underpinning is clear: GNNs are purpose-built to learn from relational data via a message-passing paradigm that explicitly models network topology. Unlike classic methods (e.g., Graph Kernels) that count predefined substructures, GNNs automatically learn the most discriminative structural motifs.

The particular strength of GAT stems from its attention mechanism. In a WebShell's call graph, not all function calls are equally important; calls to $system()$, $assert()$, or $base64\_decode()$ are far more salient than generic operations. GAT learns to assign higher attention weights to these diagnostically critical nodes and edges, effectively focusing on the parts of the call graph that best define a family's malicious signature.

\begin{table}[h!]
\centering
\caption{Optimal implementation strategies for sequence-based methods.}
\label{tab:optimal_seq}
\begin{tabular}{@{}llc@{}}
\toprule
\textbf{Method} & \textbf{Classifier} & \textbf{Optimal Strategy} \\ \midrule
\multirow{2}{*}{CBOW/GloVe} & KM/MS/RF & Avg \\
 & SVM & Concat \\ \midrule
BERT/SimCSE & All Classifiers & Concat \\ \bottomrule
\end{tabular}
\end{table}

\begin{table}[h!]
\centering
\caption{Optimal implementation strategies for graph- and tree-based methods.}
\label{tab:optimal_graph_tree}
\begin{tabular}{@{}llc@{}}
\toprule
\textbf{Method} & \textbf{Classifier} & \textbf{Optimal Strategy} \\ \midrule
\multirow{2}{*}{Graph Kernel} & Unsupervised & Subtree Kernel \\
 & Supervised & Path Kernel \\ \midrule
Tree Kernel & All Classifiers & Subtree Kernel \\ \midrule
\multirow{3}{*}{GNNs} & Unsupervised & GCN, GAT \\
 & Random Forest & GAT \\
 & SVM & GIN \\ \bottomrule
\end{tabular}
\end{table}

\subsection{Practical Implications and Guidance}

Our findings offer a clear, actionable guide for building more intelligent malware defense systems.

\paragraph{Implications for Threat Discovery and Operational Use.}
As expected, supervised classifiers achieve higher overall performance than unsupervised clustering algorithms, highlighting the value of high-quality labels for building high-precision models. Thus, when a sufficient corpus of labeled data is available, supervised classification is the preferred approach. However, in real-world security operations, labels for emerging threats are often scarce or unavailable. This is where unsupervised methods become indispensable, as their ability to group samples by intrinsic behavioral similarity provides a direct pathway for discovering new or unknown WebShell families. Our results show that in this setting, the performance gap between structural and sequential representations is magnified, making the choice of a robust structural representation even more critical. Security teams can leverage this to automatically group new malware samples, flagging emergent clusters as potential zero-day threats requiring expert analysis.

\paragraph{Optimal Implementation Strategies.}
Achieving these results requires pairing the right abstraction with the right model variant. Our benchmark provides a clear roadmap (summarized in Tables~\ref{tab:optimal_methods_by_classifier}, \ref{tab:optimal_seq}, and \ref{tab:optimal_graph_tree}).
\begin{itemize}
    \item \textbf{For overall performance}, a Tree-GAT model is the most consistent top performer across both supervised and unsupervised tasks.
    \item \textbf{For GNNs}, GAT and GCN are best for clustering, while GIN shows strength with SVMs in supervised settings.
    \item \textbf{For Graph Kernels}, Subtree Kernels are generally superior, especially for Tree Kernels where they are the optimal choice for all classifiers.
    \item \textbf{For sequence models}, the optimal aggregation strategy depends on the model architecture. For context-free embeddings like CBOW and GloVe, averaging the token embeddings of a trace is effective. For context-aware transformers like BERT and SimCSE, more sophisticated strategies like concatenating hidden states or using the final [CLS] token representation are superior.
\end{itemize}

\section{Related Work}
\label{sec:related_work}

\textbf{WebShell Detection}
Research on WebShell detection has predominantly focused on \textbf{binary classification}, distinguishing malicious from benign scripts. Early efforts relied on rule-based methods using signature matching, which proved ineffective against obfuscated or novel threats~\cite{le2021efficient, hannousse2021handling}. Subsequently, machine learning and deep learning techniques became mainstream, extracting lexical, statistical, or semantic features from source code or opcodes to train classifiers~\cite{jinping2020mixed, pu2022bert, shang2024multi, zhang2025mmfdetect}. Recently, Large Language Models (LLMs) have demonstrated strong zero-shot capabilities in this domain~\cite{han2025readthinkmitigatingllm, han2025zerotuningunlockinginitialtokens}, achieving competitive performance without task-specific fine-tuning~\cite{han2025can}.

However, while binary detection is well-studied, research on the more granular task of \textbf{WebShell family multi-classification} remains scarce. This gap is significant, as identifying the specific family of a WebShell is crucial for threat intelligence and targeted defense. A foundational contribution in this area is the MWF dataset~\cite{zhao2024malicious}, which provided the first publicly available, family-annotated dataset of malicious WebShells, thereby enabling systematic research into multi-class classification, including ours.

\textbf{Representation Learning for Program Behavior}
Our work is grounded in representation learning, which aims to transform complex, unstructured data like function call traces into meaningful vector embeddings.
Inspired by natural language processing, early methods treat program traces as sentences. Classic techniques like Word2Vec (specifically, CBOW and Skip-gram)~\cite{mikolov2013efficient} and GloVe~\cite{pennington2014glove} learn static, context-independent embeddings for each function. The advent of transformers led to powerful contextual models like BERT~\cite{devlin2018bert}, which capture deeper semantic relationships. More recently, contrastive learning methods such as SimCSE~\cite{gao2021simcse} have further improved the quality of sentence-level embeddings.

To capture the rich relational structure of function calls, we also explore graph-based methods. Traditional approaches include Graph Kernels, such as the Weisfeiler-Lehman (WL) kernel, which measure graph similarity by counting shared substructures~\cite{shervashidze2011weisfeiler}. Unsupervised methods like Graph2Vec learn embeddings for entire graphs by treating them as documents and their subgraphs as words~\cite{narayanan2017graph2vec}. 
The current state-of-the-art, however, is dominated by Graph Neural Networks, which learn node and graph representations through iterative message passing. Our work benchmarks several prominent GNN architectures: Graph Convolutional Networks ~\cite{kipf2016semi}, Graph Attention Networks~\cite{velickovic2018graph}, and Graph Isomorphism Networks~\cite{xu2018how}. 

\section{Conclusion}
\label{sec:conclusion}

In this work, we presented a systematic benchmark and comprehensive study for fine-grained WebShell family classification, a critical and underexplored task in cybersecurity. By abstracting dynamic function call traces into sequences, graphs, and trees, we conducted a large-scale evaluation of diverse representation learning methods. Our empirical results are conclusive: structural representations definitively outperform sequential models, demonstrating that a family's behavioral signature lies in its call topology, not its syntactic order. We further identified that tree-based abstractions, which preserve hierarchical execution context, provide a consistent performance advantages. Finally, we demonstrated that Graph Neural Networks, particularly GAT, are the premier architecture for this task, offering the most robust and high-performing models across both supervised and unsupervised settings.
This study moves the field beyond simple binary detection by establishing a robust baseline and providing actionable guidance for building the next generation of automated threat intelligence systems. Our findings offer a practical framework for enabling faster, more precise incident response and a more proactive defense against the evolving landscape of critical infrastructure threats.

\section*{Ethical Statement}
This research is fundamentally aimed at generating a positive societal impact by enhancing cybersecurity against malicious WebShells, a class of malware that poses a direct threat to critical infrastructure, including government, financial, and healthcare systems. The primary benefit of this work is empowering security organizations to move beyond simple detection to a more nuanced, family-level understanding of threats. This capability translates directly into tangible societal goods: it enables faster incident response to minimize data breaches of sensitive records, aids law enforcement in attributing attacks, and helps preserve the integrity and public trust in essential digital services.

A core component of our ethical methodology was the deliberate decision to release only the dynamic function call traces, not the underlying source code. This approach provides the research community with a rich behavioral summary for analysis while intentionally withholding the full, executable malicious code. By doing so, we prevent the direct redistribution or weaponization of the original malware, ensuring that our dataset serves to strengthen defenses without creating new security risks.

We acknowledge the dual-use nature of cybersecurity research, where publicizing effective methods could inform adversarial strategies. However, we contend that the net effect of this open research is overwhelmingly positive for defenders. Our focus on dynamic behavior is inherently more robust against the common obfuscation techniques used by attackers. By providing a systematic framework and sharing our findings, we aim to level the playing field, giving defenders, especially those at smaller or under-resourced organizations, the tools and knowledge to adapt more quickly. We believe the societal benefits of advancing defensive capabilities through open, responsible research significantly outweigh the inherent risks.

\bibliography{aaai2026}

\newpage
\onecolumn
\appendix
\counterwithin{figure}{section}
\counterwithin{table}{section} 

\section{Implementation Details}
\label{app: appendix_implementation}

This appendix provides detailed hyperparameter settings for our representation methods and downstream classifiers to ensure the reproducibility of our experiments. All models were implemented using Python 3.8 with PyTorch 1.12 and Scikit-learn 1.1. Experiments were conducted on a server equipped with an Intel Xeon Gold 6248R CPU, 256GB of RAM, and an NVIDIA A100 GPU.

\subsection{Hyperparameters for Representation Methods}

\paragraph{CBOW.}
We use the Word2Vec implementation from the Gensim library. The model is configured with a vector dimensionality of 128, a context window size of 5, 10 negative samples, and is trained for 100 epochs. A minimum word count of 2 was enforced. For the `concat` aggregation strategy, sequences longer than the maximum length of 256 are truncated.

\paragraph{GloVe.}
We use the official GloVe implementation. The model is configured with a context window size of 10 and an embedding dimensionality of 128. The weighting function parameter `xmax` is set to 100, and the exponent `alpha` is set to 0.75. The model was trained for 100 epochs using the Adam optimizer with a learning rate of 0.001 and a batch size of 512.

\paragraph{BERT \& SimCSE.}
We utilize the `bert-base-uncased` architecture from the Hugging Face Transformers library as the foundation for both BERT and SimCSE. The model consists of 12 transformer layers, 12 attention heads, and a hidden size of 768, which is then projected to a final embedding of 128 dimensions via a linear layer. For pre-training, we construct a domain-specific corpus where each line contains two randomly selected function call sequences. The model is trained for 10 epochs with a batch size of 64, a learning rate of 2e-5, and the AdamW optimizer. For SimCSE, we use a dropout rate of 0.1 as the noise operator for the contrastive learning objective.

\paragraph{Graph Kernels.}
We use the `gklearn` library. For the Weisfeiler-Lehman (WL) subtree kernel, the number of iterations was set to 5. For the Random Walk (RW) kernel, the random walk length was set to 10.

\paragraph{Graph Edit Distance (GED).}
Our GED computation is implemented using the `gklearn` library with the following settings:
\begin{itemize}
    \item \textbf{Edit Costs:} We define a constant edit cost vector of `[1, 1, 1, 1, 1, 1]` for node/edge deletion, insertion, and substitution. This uniform cost treats all structural changes equally.
    \item \textbf{GED Algorithm:} We use the \texttt{BIPARTITE} graph matching algorithm for its efficiency on large-scale graph data.
\end{itemize}

\paragraph{Graph2Vec.}
We use the official implementation of Graph2Vec. The model is configured with an embedding dimensionality of 128, 10 negative samples, and a WL subtree height of 3. It was trained for 100 epochs with a learning rate of 0.025.

\paragraph{Graph Neural Networks (GNNs).}
All GNN models (GCN, GAT, GIN) were implemented using PyTorch Geometric. Each model consists of 3 GNN layers followed by a global mean pooling layer and a 2-layer MLP head to produce the final 128-dimensional embedding. We trained for 200 epochs using the Adam optimizer with a learning rate of 0.001 and a batch size of 64. A dropout rate of 0.5 was applied after each GNN layer. For GAT, we used 4 attention heads.

\subsection{Hyperparameters for Downstream Classifiers}
To ensure a fair and robust comparison, we used a fixed set of optimized hyperparameters for our downstream classifiers, identified via grid search on a validation set.

\paragraph{K-Means.}
The number of clusters (`k`) is set to the ground-truth number of families in each dataset. We used the "k-means++" initialization method and set `n\_init` to 10 to ensure stability.

\paragraph{Mean-Shift.}
The bandwidth parameter was automatically estimated using the `estimate\_bandwidth` function from Scikit-learn on a sample of the data.

\paragraph{Random Forest.}
The model is configured with 100 estimators (trees), a maximum depth of 10, and a minimum of 5 samples required to split an internal node.

\paragraph{Support Vector Machine (SVM).}
We use an SVM with a radial basis function (RBF) kernel. The regularization parameter `C` is set to 1.0, and the kernel coefficient `gamma` is set to `scale`.

\newpage
\section{Prompt Templates for LLM-Powered Data Augmentation}
\label{app: appendix_prompts}

\subsection{Prompt for Intra-Family Augmentation}
The following prompt was used to generate new, diverse samples for existing WebShell families. The goal was to create traces that are behaviorally consistent with the target family while introducing syntactic variations.

\begin{tcolorbox}[title={Prompt for Intra-Family Data Augmentation}, colframe=blue!50!black, colback=blue!10!white, top=1mm, bottom=1mm]
\textbf{System Prompt:} You are a cybersecurity expert specializing in malware analysis. Your task is to generate new, plausible dynamic function call traces for a specific WebShell family. The generated traces must be behaviorally consistent with the provided description and examples, but should introduce minor variations to enhance data diversity.

\textbf{User Prompt:} Based on the following behavioral profile and examples for the \textbf{[Family Name]} WebShell family, please generate 10 new and unique dynamic function call traces.

\textbf{[Behavioral Description]}

\textit{For Example: This family typically uses base64 decoding on POST data and then executes the result using an `eval` or `assert` call. It often includes file manipulation functions like `fopen` and `fwrite` for persistence.}

\textbf{[Example Traces]}

\textit{
For Example:\\
1. ["\_main\_", "base64\_decode", "eval", "zend\_fetch\_r\_post", ...]\\
2. ["\_main\_", "zend\_compile\_string", "assert", "base64\_decode", ...]}

\textbf{Output:}
\end{tcolorbox}

\subsection{Prompt for New Family Simulation}
This prompt was designed to simulate adversarial innovation by instructing the LLM to create a novel WebShell family. This is achieved by blending the characteristics of two existing families, thereby generating data for zero-day threat scenarios.

\begin{tcolorbox}[title={Prompt for New Family \& Zero-Day Simulation}, colframe=red!60!black, colback=red!10!white, top=1mm, bottom=1mm]
\textbf{System Prompt:} You are an expert malware author. Your objective is to design a novel WebShell family by creatively blending the characteristics of two existing malware families. First, describe the core behavior and tactics of your new creation. Then, generate function call traces that reflect this new, hybrid behavior.

\textbf{User Prompt:} Design a new WebShell family that combines the stealth techniques of \textbf{[Family A Name]} with the command execution capabilities of \textbf{[Family B Name]}.
\begin{enumerate}
    \item Provide a short description of the new family's behavior.
    \item Generate 10 dynamic function call traces for this new family.
\end{enumerate}

\textbf{[Family A Profile: Name and Behavioral Description]}\\
\textit{e.g., Family A (Stealthy Dropper): Focuses on obfuscation using string manipulation functions and avoids direct execution calls. It writes payloads to temporary files.}

\textbf{[Family B Profile: Name and Behavioral Description]}\\
\textit{e.g., Family B (Powerful C2): Uses direct command execution via `shell\_exec` and `system` and communicates over raw sockets.}

\textbf{[Example Traces from Family A \& B]}

\textbf{Output:}
\end{tcolorbox}

\newpage
\section{Results for DS1, DS2, and DS3}
\label{app: ds1_2_results}

\begin{figure*}[h!]
    \centering
    \includegraphics[width=1\linewidth]{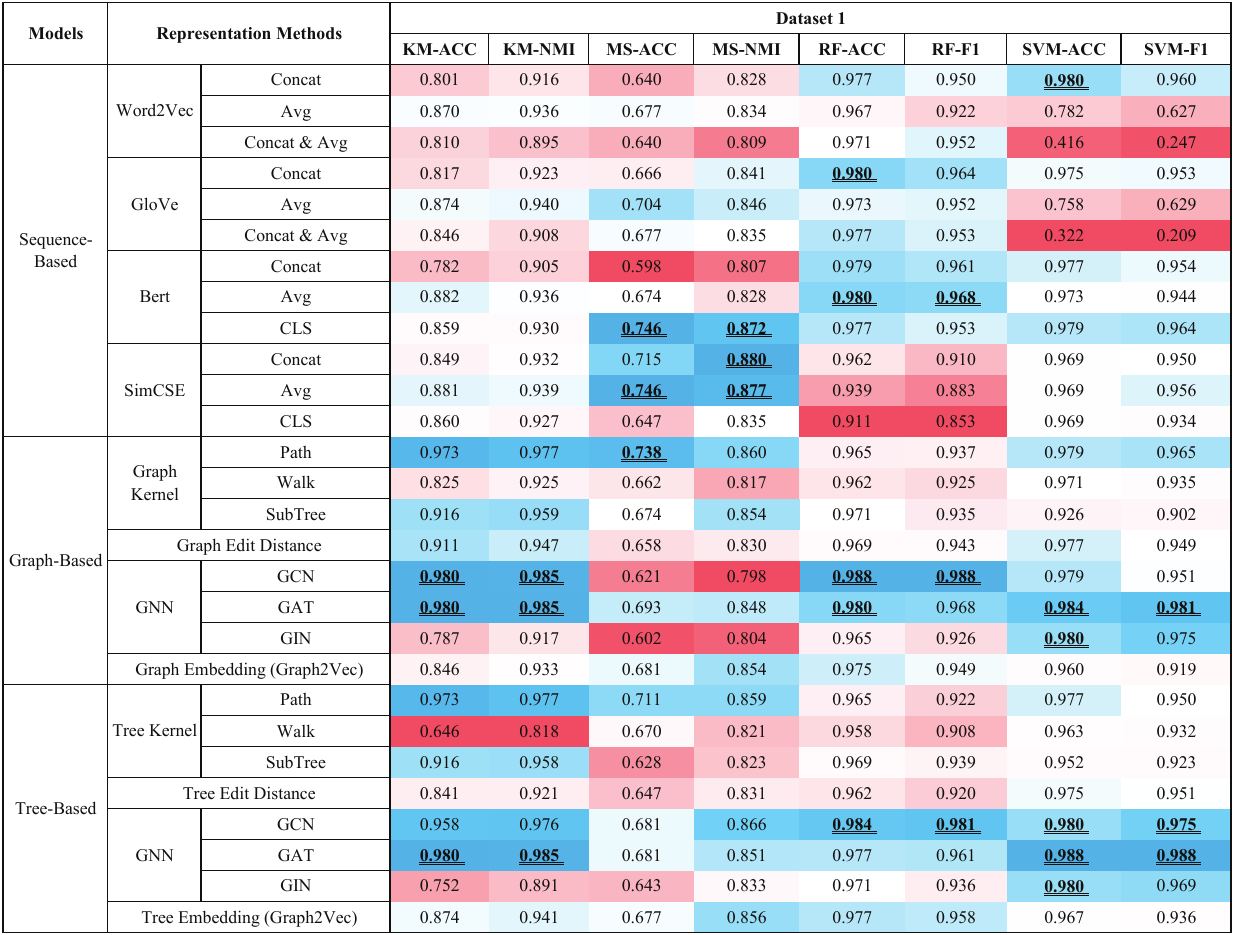}
    \caption{Performance comparison of all representation methods on the DS1 dataset.}
    \label{fig:ds1}
\end{figure*}

\begin{figure*}[h!]
    \centering
    \includegraphics[width=1\linewidth]{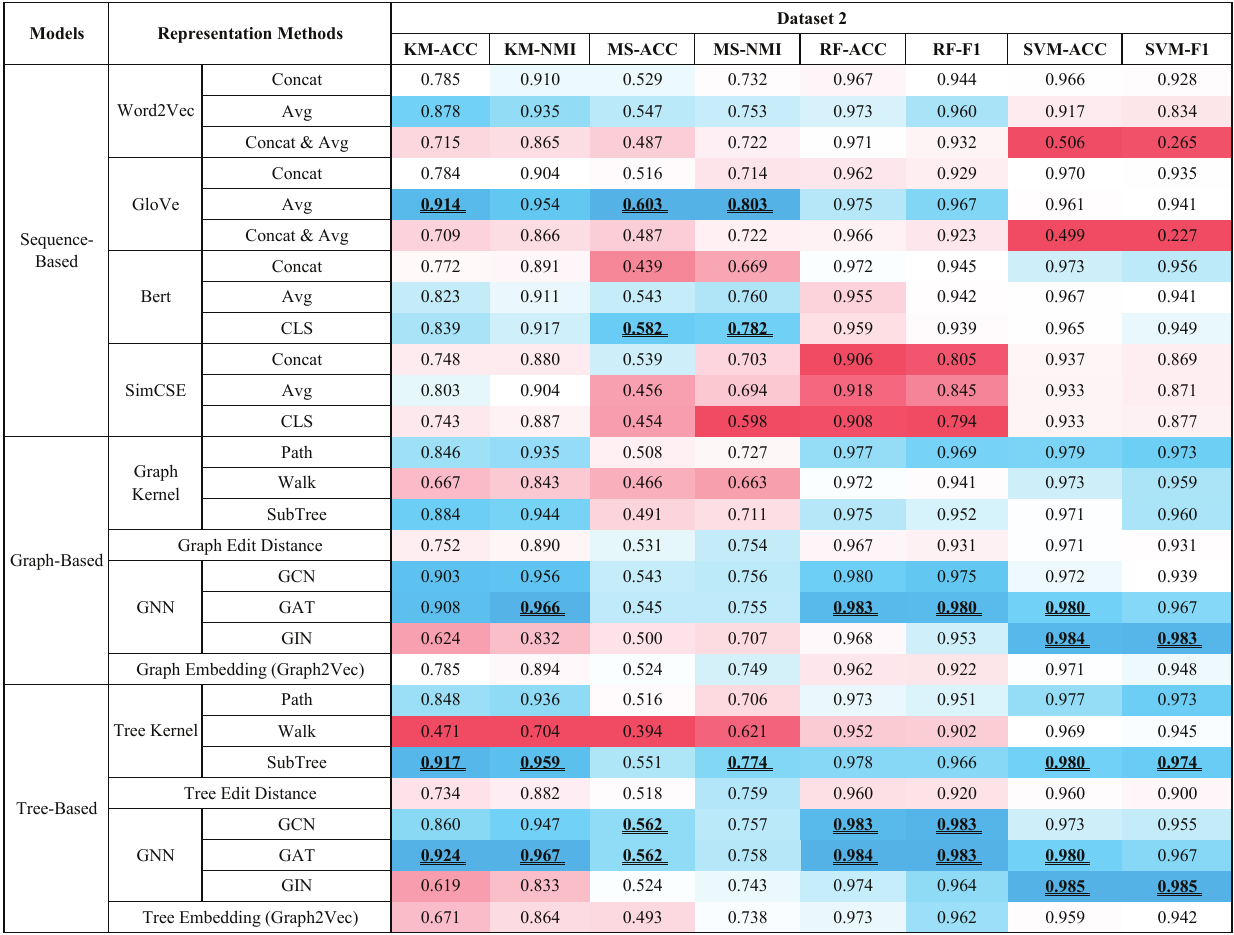}
    \caption{Performance comparison of all representation methods on the DS2 dataset.}
    \label{fig:ds2}
\end{figure*}

\begin{figure*}[h!]
    \centering
    \includegraphics[width=1\linewidth]{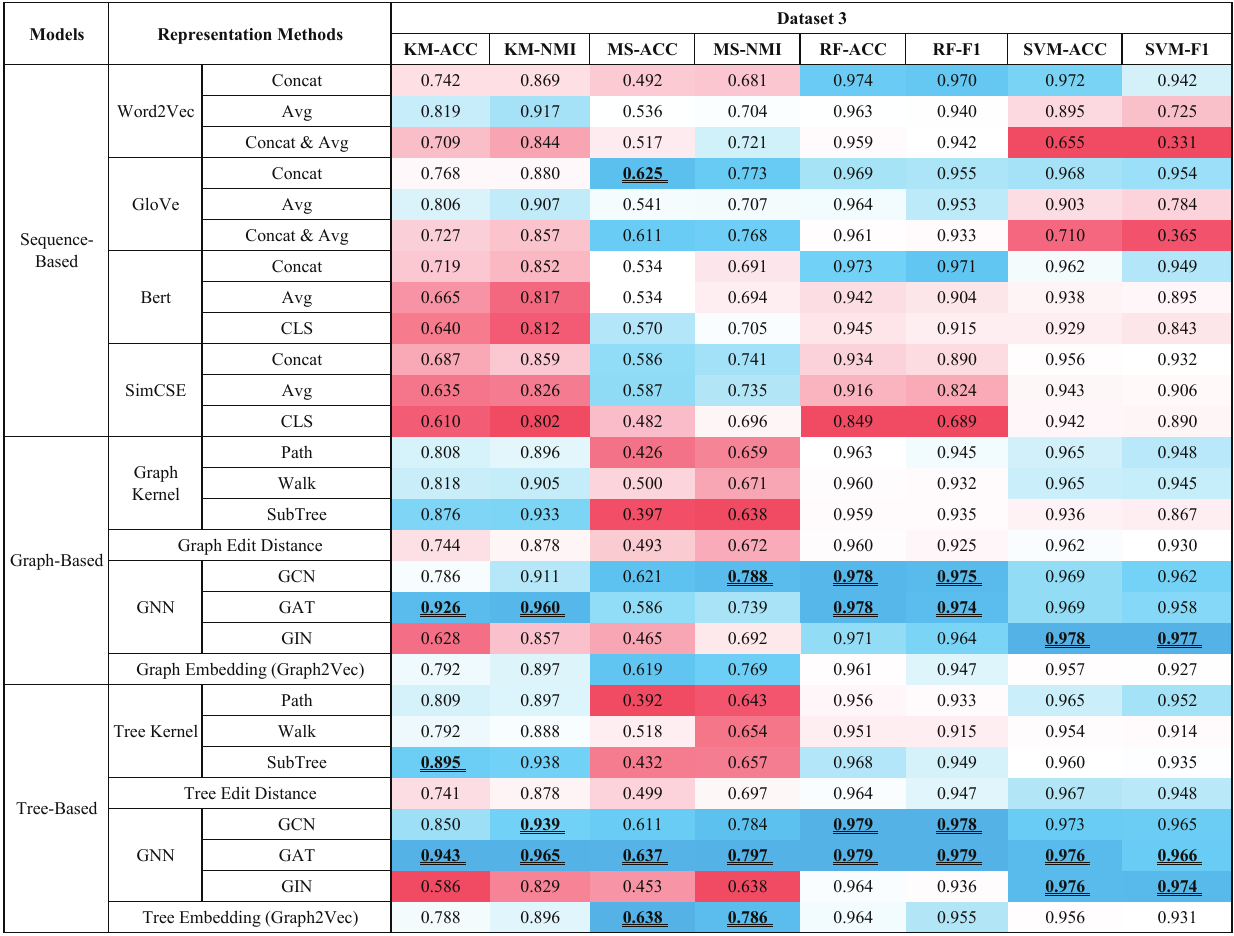}
    \caption{Performance comparison of representation methods on the DS3 dataset.}
    \label{fig:ds3}
\end{figure*}

\end{document}